\documentclass{article}

\PassOptionsToPackage{numbers}{natbib}

\usepackage[preprint]{neurips_2025}

\usepackage{enumitem}
\setlist[itemize]{noitemsep}

\usepackage[utf8]{inputenc} 
\usepackage[T1]{fontenc}    
\usepackage{hyperref}       
\usepackage{url}            
\usepackage{booktabs}       
\usepackage{amsfonts}       
\usepackage{nicefrac}       
\usepackage{microtype}      
\usepackage{xcolor}         
\usepackage{amsmath}

\usepackage{siunitx}

\title{GeoGraph: Geometric and Graph-based Ensemble Descriptors for Intrinsically Disordered Proteins}

\author{%
  Eoin Quinn \\
  InstaDeep \\
  London, UK \\
  \texttt{e.quinn@instadeep.com} \\
  \And
  Marco Carobene \\
  InstaDeep \\
  Berlin, Germany \\
  \texttt{m.carobene@instadeep.com} \\
  \And
  Jean Quentin \\
  InstaDeep \\
  Paris, France \\
  \texttt{j.quentin@instadeep.com} \\
  \And
  Sebastien Boyer \\
  InstaDeep \\
  Paris, France \\
  \texttt{s.boyer@instadeep.com} \\
  \And
  Miguel Arbes\'u \\
  InstaDeep \\
  Berlin, Germany \\
  \texttt{m.arbesu@instadeep.com} \\
  \And
  Oliver Bent \\
  InstaDeep \\
  Paris, France \\
  \texttt{o.bent@instadeep.com}  
}

\begin{document}

\maketitle

\begin{abstract}
While deep learning has revolutionized the prediction of rigid protein structures, modelling the conformational ensembles of Intrinsically Disordered Proteins (IDPs) remains a key frontier. Current AI paradigms present a trade-off: Protein Language Models (PLMs) capture evolutionary statistics but lack explicit physical grounding, while generative models trained to model full ensembles are computationally expensive. In this work we critically assess these limits and propose a path forward. We introduce GeoGraph, a simulation-informed surrogate trained to predict ensemble-averaged statistics of residue–residue contact-map topology directly from sequence. By featurizing coarse-grained molecular dynamics simulations into residue- and sequence-level graph descriptors, we create a robust and information-rich learning target. Our evaluation demonstrates that this approach yields representations that are more predictive of key biophysical properties than existing methods. 

\end{abstract}

\section{Introduction}

Proteins are the cell’s molecular machines: sequence-encoded biopolymers which catalyze reactions, regulate processes, and shape cellular architecture. Recent years have witnessed a paradigm shift in protein modelling, driven by advances in experimental techniques and the maturation of deep learning.
In particular, the rapid growth of high-throughput sequencing has been pivotal \cite{theuniprotconsortium2025}. On the one hand it has  enabled language-modelling approaches, especially Masked Language Modelling (MLM), to learn the statistical patterns of evolution directly from vast, unannotated sequence databases \cite{Rives2021,Lin2023ESMFold}. On the other, Multiple Sequence Alignments (MSAs), coupled with decades of structure determination experiments \cite{Berman2000}, underpin deep learning models like AlphaFold \cite{Jumper2021} and RosettaFold \cite{baek2021rosettafold}, which now achieve near-experimental accuracy for a broad class of structured proteins.

With static structures largely tractable, the frontier of computational structural biology is advancing toward a more fundamental problem: modelling the full conformational ensemble -- the Boltzmann distribution of conformations under physiological solution conditions. To frame this challenge, we can identify three regimes along the structural order-disorder continuum: (i) proteins that adopt a single, highly stable fold; (ii) dynamic proteins that interconvert among a few metastable states; and (iii) Intrinsically Disordered Proteins (IDPs), which manifest a broad, heterogeneous set of rapidly fluctuating conformations \cite{DysonWright2005,VanderLee2014}. The first regime is where models trained on protein crystal structures excel. The second is well-captured by Markov State Models (MSM), which characterise the ensemble by the populations of metastable states and the kinetic rates between them, typically inferred from long Molecular Dynamics (MD) simulations \cite{Prinz2011,Chodera2014,Husic2018}. The third regime of IDPs is, however, particularly challenging, and provides the focus for this work. Beyond the inherent complexity of modelling a heterogeneous ensemble, these proteins also face significant experimental and evolutionary hurdles. Experimentally, obtaining data is laborious, and their dynamic nature means measurements are typically averaged across the entire ensemble and/or over time. Evolutionarily, they exhibit poor sequence conservation, a characteristic thought to derive from the lack of a stable structure required to maintain function \cite{brown2011idpevolution}.

A recent line of work aims to use deep generative models, especially diffusion models, to map sequence directly to a full conformational distribution \cite{lewis2025bioemu, jin2025p2dflow, janson2024idpsam, zhu2024idpfold}. While useful, this strategy faces practical and statistical hurdles: generating, storing, and analyzing thousands of conformers per protein is expensive, and for many downstream tasks such high-dimensional stochastic detail can obscure the underlying biophysical signal. From a statistical physics perspective, fluctuations faster than the timescale of interest are effectively marginalized as entropy, making the explicit modelling of  fine-grained, high-frequency detail counterproductive. Indeed timescale separation underpins MSM coarse-graining, which emphasizes slow, kinetically relevant transitions between metastable states, rather than the noisy internal motions within them \cite{Chodera2014,Husic2018}.

Here we take a different approach: rather than modelling entire ensembles explicitly, we model their aggregate properties directly. Specifically, we propose to extract essential biophysical content of an IDP ensemble from the statistics of its transient residue–residue contacts \cite{brinda2005, amitai2004}. The power of this approach has recently been demonstrated by WARIO \cite{gonzalezdelgado2024wario}, which uses contact-based descriptors to cluster simulation trajectories of IDPs into structurally coherent states. Our work leverages this same insight for a different purpose: instead of post-hoc analysis of a single ensemble, our aim is high-throughput prediction directly from sequence. To achieve this, we convert conformations from simulation into residue-level contact-map graphs, compute a diverse set of graph-theoretic descriptors, and use their ensemble-averaged values as the direct prediction targets for our model. This approach acts as a deliberate information bottleneck, filtering high-frequency fluctuations while preserving the stable signature of the ensemble.

A key design choice is resolution. We operate at the residue level--a natural middle ground between whole-sequence and all-atom representations. Unlike models that predict a few global aggregates and lose positional detail, we learn a rich vector of aggregate properties per residue, capturing biophysical characteristics across the protein sequence.

We make our code, pre-trained models and datasets publicly available at \url{https://github.com/instadeepai/GeoGraph}.

\section{Related work}

Our work builds on several research threads at the intersection of machine learning and protein science. One major line of work uses deep generative models to sample full conformational ensembles from sequence, for both general proteins \cite{lewis2025bioemu, jin2025p2dflow, daigavane2025jamun} and specifically for IDPs \cite{novak2025, janson2023idpgan, janson2024idpsam, zhu2024idpfold}. While powerful, these methods can be computationally expensive, and general-purpose models often rely on co-evolutionary signals from MSAs that are absent in IDPs. An alternative approach, more aligned with our own, is to predict ensemble-averaged aggregate properties directly from sequence. For example, ALBATROSS \cite{lotthammer2024} predicts five global geometric properties of IDPs, while IDP-BERT \cite{mollaei2024idpbert} fine-tunes a protein language model for similar tasks. We extend this strategy by introducing GeoGraph, a model that learns a richer representation by predicting a diverse set of residue-level geometric and graph-theoretic descriptors. This method is inspired by the long history of using residue contact networks to analyze protein structure and stability \cite{brinda2005, amitai2004} and is complementary to the recent method WARIO \cite{gonzalezdelgado2024wario} which uses contact maps for post-hoc characterization of individual IDP ensembles.

\section{GeoGraph}

Our goal is to learn residue-level representations of IDPs from molecular dynamics (MD) simulations to capture essential physical principles missed by protein language models (PLMs) and methods trained on static, folded proteins.

Generating the vast simulation data required for deep learning is computationally prohibitive with high-fidelity all-atom force fields.
We therefore use CALVADOS-2 \cite{Tesei2023CALVADOS2}, a state-of-the-art one-bead-per-residue coarse-grained force field designed for IDPs, and experimentally validated on SAXS and FRET measurements. While this approach sacrifices fine-grained detail, its design is based on an effective description of non-bonded interactions, which enables it to excel at capturing transient residue-residue contact patterns.

We hypothesise that these transient contacts encode rich physicochemical information, which we formalize by analyzing their aggregate properties. For each conformation, we construct a residue-contact graph (8\AA~cutoff), compute a diverse set of node- and graph-level features, and average these across the full ensemble. This yields  a stable statistical fingerprint of the protein’s dynamic structure which serves as a direct prediction target.

Our model, GeoGraph, employs a sequence-to-sequence architecture with a 4-layer transformer encoder backbone ($\sim$ 2M parameters) that maps an amino acid sequence to residue-level embeddings. These embeddings are fed as input to separate shallow MLP heads to predict properties at both the sequence- and residue-level. For full details see Appendix \ref{app:model}.

We consider two flavours of descriptors, which we refer to as \textit{geometric} and \textit{graph}-based. The geometric descriptors are commonly-employed sequence-level measures of IDP conformational ensembles: end-to-end distance ($R_e$),  radius of gyration ($R_g$), asphericity ($\Delta$), and the Flory scaling exponent ($\nu$) and  prefactor ($A_0$). Both $R_g$ and $R_e$ can be experimentally determined, and in turn used to determine the Flory prefactor and exponent \cite{ahmed2019}. Small Angle X-ray Scattering (SAXS) yields the ensemble-averaged radius of gyration $\langle R_g \rangle$, whereas Fluorescence resonance energy transfer (FRET) spectroscopy yields $\langle R_e \rangle$, or even $R_e$ distributions in the case of single-molecule FRET \cite{hofmann2012}. 

For the graph-based descriptors we consider both sequence- and residue-level features, which capture diverse properties such as network compactness (global efficiency), mixing patterns (assortativity), and residue importance (degree and betweenness centrality), providing a rich, physics-informed learning signal. For full details see Appendix \ref{app:features}.

We consider multiple variants of GeoGraph so as to clearly dissect its behaviour. Our main model is:\vspace{-0.5em}
\begin{itemize}
    \item {\bf GeoGraph:} the full architecture described above, containing the transformer backbone with both sequence-level and residue-level prediction heads, and trained end-to-end to predict the full suite of geometric and graph-based features.
\end{itemize}
We complement this with baseline variants as follows:
\vspace{-0.5em}
\begin{itemize}
\item {\bf Geo:} a baseline model trained to predict only the sequence-level geometric features, i.e.~those used as benchmarks. I.e.~the prediction of the graph-based features is omitted from the training. This serves as an analogue to ALBATROSS, up to the change in architecture and the use of a single model to predict all features.
\item {\bf Geo-zero:} a greatly simplified variant of the Geo model where the transformer backbone has zero layers. This tests the performance of contextless token embeddings, and provides a naive minimal performance floor.
\item {\bf Graph:} a variant of GeoGraph designed to assess the transferability of the learned embeddings. Trained in two stages: first, the full model is trained end-to-end to predict only the graph-based features; second, the backbone weights are frozen, and a new sequence-level prediction head (GeoHead) is trained to predict the geometric features from the learned embeddings.
\end{itemize}

We train and evaluate our models on the Human--IDRome dataset \cite{tesei2024idrome}, containing simulated conformational ensembles for 28,058 intrinsically disordered regions from the human proteome. This is the largest publicly available dataset of its kind, which makes it ideal for benchmarking differing approaches. The ensembles were generated using the CALVADOS-2 coarse-grained force field, with each sequence represented by 1,000 weakly correlated conformational frames sampled from the simulation trajectory \cite{tesei2024idrome}. We partition the dataset using a 80/10/10 split based on sequence similarity, see Appendix \ref{app:split}.

\section{Evaluation}\label{sec:evaluation}
To benchmark performance we evaluate models on their ability to predict the five geometric features $(R_e, R_g, \Delta, \nu, A_0)$, which are well-studied, experimentally relevant measures of IDP conformational ensembles. Results are presented in Table~\ref{table:main}, and ablation on the GeoGraph model is provided in Table~\ref{table:ablation}. We highlight that while scores for $R_e$ and $R_g$ are high across the board, the true test is the performance on the more complex shape descriptors $(\Delta, \nu, A_0)$, on which GeoGraph excels.

\begin{table}[t]
\centering
\begin{tabular}{l | l l l l l} 
 \hline
 
  & $R_{e}$ & $R_{g}$ & $\Delta$  & $\nu$ & $A_0$ \\ [0.2ex] 

  \hline

{\bf GeoGraph} & {\bf 0.993} (0) & {\bf 0.996} (0) & {\bf 0.899} (5) 
& {\bf 0.893} (6) & {\bf 0.875} (16) \\

 \hline

Geo & 0.991 (2) & 0.994 (1) & 0.875 (13) 
& 0.856 (14) & 0.787 (30) \\ 

Geo-zero & 0.596 (33) & 0.603 (33) & 0.584 (6) & 0.505 (7) & 0.389 (13) \\

Graph $\to$ GeoHead & 0.992 (1) & {\bf 0.996} (0) & 0.864 (13) 
& 0.854 (15) & 0.793 (32) \\

 \hline

 STARLING & 0.914 & 0.951 & -0.460 
 & 0.261 & 0.386 \\
 
 STARLING (retrained) & 0.983 & 0.992 & 0.314 
 & 0.677 & 0.539 \\
  
 \hline
 
  ALBATROSS & 0.899 & 0.932 & 0.441 & 
  0.275* & -0.471*\\ 

 

  ALBATROSS (retrained) & 0.970 & 0.984 & 0.790 & 
  0.698 & 0.513 \\
 
 \hline

 ESM2-8M & 0.983 (1) & 0.991 (1) & 0.754 (8) 
 & 0.684 (8) & 0.523 (19) \\ 

 IDP-ESM2-8M & 0.982 (1) & 0.987 (1) & 0.783 (2) 
 & 0.767 (5) & 0.643 (14) \\ 
 
 ESM2-150M & 0.984 (1) & 0.991 (1) & 0.792 (2) 
 & 0.763 (4) & 0.637 (5) \\ 

 IDP-ESM2-150M & 0.980 (1) & 0.986 (1) & 0.786 (6) 
 & 0.777 (4) & 0.660 (7) \\ 

 \hline
 
 \end{tabular}
\caption{$R^2$ scores for the IDP property prediction task on our Human--IDRome test set. Where parentheses are shown, the results are the mean of 5 models with different random seeds, along with the standard error on the final digits.  We highlight with (*) that the $R^2$ scores of the pretrained ALBATROSS models for $\nu$ and $A_0$ may be affected by differences in computation of the scaling parameters between our work and theirs (see Appendix~\ref{app:albatross_flory}).}
\label{table:main}
\end{table}

We compare against two leading IDP methods: STARLING \cite{novak2025}, a generative diffusion model, and ALBATROSS \cite{lotthammer2024}, an RNN-based direct predictor. Since the original models were trained on data generated with a different force field, we retrained them on our dataset for a fair comparison. See Appendix~\ref{app:IDP} for further details.

In their analysis of the Human–IDRome dataset \cite{tesei2024idrome}, the authors trained a support vector regression model on a manually chosen set of sequence features for modelling the Flory scaling exponent $\nu$. For comparison we evaluated this on our test set, and find it achieves an $R^2$ score of 0.765. The author's also quantified the simulation (convergence) error for $\nu$ as $\sim 0.01$. For GeoGraph we find the corresponding RSME as 0.011, which provides a strong sanity check that the model is learning the physical signal right up to the limit of the data's precision.

We next compare the sequence-to-sequence backbone against Protein Language Model (PLM) embeddings. We use ESM-2 \cite{Lin2023ESMFold} and test the model in two settings. Firstly, we used the general-purpose pre-trained embeddings of the 8M and 150M models. Secondly, we curated a dataset of 30 million IDP sequences, which we refer to as \textit{IDP-Euka-90}, and used this to fine-tune two corresponding versions of ESM-2, \textit{IDP-ESM2-8M} and \textit{IDP-ESM2-150M}, see Appendix~\ref{app:IDP-ESM} for further details. In both cases, we freeze the backbone model and train a sequence-level prediction head for predicting the geometric features -- as we did for the Graph model above.

Finally, we attempted to evaluate BioEmu, a large-scale general-purpose ensemble emulator \cite{lewis2025bioemu} which uses a diffusion model to generate conformational ensembles conditioned on the MSA of a sequence. Due to computational constraints, we were not able to generate sufficiently large ensembles with BioEmu on our test set to make a fair comparison. In a small experiment where we generated 1000 conformers/sequence for 100 randomly-sampled test sequences, we observed very poor performance ($R^2 < 0$ for all features), which is consistent with recent work evaluating BioEmu for IDPs \cite{schnapka2025atomic}, and may be explained by the poor sequence conservation of IDPs.

\begin{table}[b]
\centering
\begin{tabular}{l | l l l l l l} 
 \hline
 
  & $R_{e}$ & $R_{g}$ & $\Delta$  
  & $\nu$ & $A_0$ \\ [0.5ex] 
  
 \hline

GeoGraph (4 layers) & {\bf 0.993} (0) & {\bf 0.996} (0) & {\bf 0.899} (5) & {\bf 0.893} (6) & {\bf 0.875} (16) \\
-- 6 layers & {\bf 0.993} (1) & {\bf 0.996} (1) & {\bf 0.897} (3) & {\bf 0.891} (4) & {\bf 0.872} (15) \\
-- 2 layers & {\bf 0.992} (1) & {\bf 0.996} (0) & 0.890 (6) & 0.883 (3) & 0.848 (10) \\
-- 1 layer & {\bf 0.991} (2) & {\bf 0.994} (2) & 0.864 (26) & 0.859 (14) & 0.794 (31) \\
-- w/o sequence graph features & {\bf 0.993} (1) & {\bf 0.996} (1) & 0.896 (9) & 0.886 (9) & 0.858 (15) \\
-- w/o residue graph features & 0.988 (2) & 0.992 (2) & 0.858 (10) & 0.856 (15) & 0.806 (34) \\
-- w/o residue centralities & {\bf 0.993} (1) & {\bf 0.996} (0) & 0.886 (8) & 0.880 (12) & 0.848 (26) \\ 
 -- w/o residue pagerank & {\bf 0.993} (1) & {\bf 0.996} (1) & {\bf 0.896} (10) & {\bf 0.889} (7) & {\bf 0.868} (18) \\ 
-- w/o residue clustering & {\bf 0.993} (1) & {\bf 0.996} (0) & 0.897 (4) & 0.886 (6) & 0.861 (16) \\ 

 \hline
 
 \end{tabular}
\caption{$R^2$ scores on our Human--IDRome test set for several ablations on the GeoGraph model. The results are the mean of 5 models with different random seeds, along with the standard error on the final digits in parentheses. For each task the best performing results (within error) are in bold.}
\label{table:ablation}
\end{table}

\section{Discussion}\label{sec:discussion}

As shown in  Table~\ref{table:main}, GeoGraph achieves highly competitive performance against leading methods for IDP ensemble property prediction, in particular on the more complex shape descriptors $(\Delta, \nu, A_0)$. 
Critically, our model predicts these descriptors several orders of magnitude faster than it takes to run the CALVADOS-2 simulator that it emulates: GeoGraph can process the entire test set of 2,388 sequences in approximately 1 second on a single GPU (H100), whereas simulation of these ensembles on Google Colab takes on the order of 10 days \cite{tesei2024idrome}.

The source of GeoGraph's strong performance can be seen by comparing our model variants: the Geo model, which trains on geometric features alone, and the Graph model, which learns representations solely from graph topology (and is evaluated on the GeoHead trained on these). Both variants perform on par with each other, demonstrating that the rich biophysical information in the contact-map topology is sufficient to create representations as powerful as those learned by direct optimization. The value of these learned representations is confirmed by the far superior performance of Graph relative to our Geo-zero baseline, which lacks this contextual learning. Crucially, the main GeoGraph model outperforms both specialized variants, demonstrating a clear synergistic effect. This supports our central hypothesis: the auxiliary task of predicting contact map characteristics is a highly beneficial component for extracting transferable representations from MD simulation data. Furthermore, the ablation in Table~\ref{table:ablation} reveals that the context-aware, residue-level graph features are the primary drivers of this learning.

When compared to the generative baseline of STARLING, our direct-prediction approach appears more robust for capturing complex shape descriptors. This suggests that inferring these aggregate properties from a generated ensemble can be a less effective approach. The improved performance over ALBATROSS, an analogous direct predictor, can be attributed to our model's larger capacity and richer, residue-level feature set. 

For the comparison against PLM embeddings, the most direct reference point is with our Graph $\to$ GeoHead model.  We see that our simulation-informed embeddings provide a significantly stronger predictive signal for geometric properties, even after fine-tuning ESM-2 on a large corpus of IDP sequences.  While a superior performance may be expected when the training objective aligns with the evaluation task, the magnitude of the difference underscores a key limitation of protein language models when applied to IDPs: their reliance on evolutionary patterns, which serve as a noisy and incomplete proxy for the physical properties that govern dynamic ensembles \cite{brown2011idpevolution}.

We observe that the IDP fine-tuning of ESM-2 leads to a significant performance improvement for the 8M model, while having little effect on the 150M model. We hypothesize that  ESM2-150M already captures key properties of IDP sequences from its UniRef50 pretraining, and that additional fine-tuning does not significantly enhance its ability to model geometric features.

While the results of our evaluation are highly encouraging, this exploratory study has several key limitations. GeoGraph is fundamentally an emulator of the CALVADOS-2 coarse-grained simulation, inheriting its lack of all-atom detail. Furthermore, our contact map featurization is simplistic, and by predicting only the mean of each descriptor, we lose valuable information about the ensemble's heterogeneity. Future work could directly address these points by training on all-atom data and enriching the prediction targets to include higher-order statistics like variance. 
A next iteration of the GeoGraph approach could learn an optimal graph construction and its most relevant features directly from the data.

\section{Conclusion}\label{sec:conclusion}

In this work we introduce GeoGraph, a sequence-to-sequence model trained to predict aggregate properties of IDP conformational ensembles. It achieves this by first featurizing individual conformations from MD simulations into contact-graph topologies, and then learning to predict the ensemble average of these features, at both a residue- and sequence-level. Our evaluation demonstrates that this approach not only achieves highly competitive performance on benchmark tasks but also yields embeddings that are more effective for predicting key experimentally relevant properties than existing methods. Our trained GeoGraph and IDP-ESM2 models, along with the IDP-Euka-90 training dataset are available at \url{https://github.com/instadeepai/GeoGraph}.


\bibliographystyle{plain} 
\bibliography{references}

\appendix

\section{Additional details}\label{app:model}

\subsection{GeoGraph}

GeoGraph is a sequence-to-sequence model that maps a protein’s amino-acid sequence to feature vectors describing aggregate physical properties at both the sequence- and residue-level. The backbone is a transformer encoder \cite{vaswani2017attention}, chosen for its ability to capture long-range dependencies and produce context-rich embeddings. We build on the Hugging Face implementation of ESM-2 \cite{Lin2023ESMFold,Wolf2020Transformers}, which uses Pre-Layer Normalization (Pre-LN) \cite{Xiong2020PreLN} and Rotary Position Embeddings (RoPE) \cite{Su2021RoPE}.

We use a 4-layer transformer with hidden size 256, 4 attention heads, and a feed-forward expansion factor of 2 (FFN dimension 512), for a total of $\approx$ 2.2M parameters. The output of the transformer is a sequence of residue-level embeddings. We obtain a single sequence-level embedding by taking the mean of these residue-level embeddings, a simple yet robust method for creating a global representation.

To predict targets, we attach separate heads for sequence-level and residue-level features. Each head is a shallow MLP with a single hidden layer of dimension 128 and a dropout probability of 0.1, so that performance primarily reflects the backbone’s context-aware embeddings.

To ensure robust training, the transformer backbone is also regularized with dropout of 0.1 on both the FFN activations and the attention probabilities. We use the Adam optimizer and a cosine learning rate scheduler with warmup, with peak a learning rate   of 5e-4, and a batch size of 512.

For the Graph model, where we train the prediction head for the geometric features in a second stage on a frozen backbone, we used the same batch size with a peak learning rate of 3e-3.

\subsection{Features}\label{app:features}

We consider two flavours of descriptors, which we refer to as geometric and graph-based. The geometric descriptors all sequence-level features, while for the graph-based descriptors we consider both sequence- and residue-level features.
Graph features are computed using python's NetworkX package (default settings) \cite{hagberg2008},
and for training our models we standardise all target features to have zero mean and unit variance.

\noindent
{\bf Geometric, sequence-level:}
We consider commonly employed measures of IDP conformational ensembles of computable from MD simulation frames:
{\bf end-to-end distance} ($R_e$),   {\bf radius of gyration} ($R_g$), {\bf asphericity} ($\Delta$), and the Flory scaling {\bf exponent} ($\nu$) and {\bf prefactor} ($A_0$).  

\noindent
{\bf Graph, sequence-level:}
As not all graphs were connected we computed {\bf fragmentation index} as the fraction of nodes in the Largest Connected Component (LCC); {\bf average shortest path length} on the LCC and {\bf global efficiency} on the full graph to quantify compactness/communication; {\bf average clustering} and {\bf transitivity} as measures of local triadic closure; and {\bf degree assortativity} as well as {\bf charge assortativity} and {\bf hydrophobicity assortativity} to assess mixing patterns. 

\noindent
{\bf Graph, residue-level:}
Here we included {\bf degree centrality} (local contact density), {\bf betweenness centrality} (bridging propensity), {\bf harmonic centrality} (inverse-distance reachability), {\bf PageRank} \cite{brin1998pagerank}, {\bf core number}, {\bf local clustering coefficient}, and as well as an {\bf in-largest-connected-component} indicator. 

\subsection{Human–IDRome dataset}\label{app:split}

We partitioned the Human--IDRome dataset \cite{tesei2024idrome} based on sequence similarity into 80/10/10 splits for training, validation, and testing. To ensure fair comparison with prior work, this split was performed using MMseqs2 \cite{steinegger2017mmseqs2} with parameters (min\_seq\_id=0.7, coverage=0.8, cov\_mode=1), identical to the parameters used by STARLING \cite{novak2025}. Additionally, we filtered the dataset to sequences with a maximum length of 256 residues.

\section{IDP-ESM}\label{app:IDP-ESM}

For training our fine-tuned versions of ESM-2, \textit{IDP-ESM2-8M} and \textit{IDP-ESM2-150M}, we  curated a large dataset of biological IDP sequences, which we call \textit{IDP-Euka-90}.  As suggested in the Metapredict V3 paper \cite{Lotthammer2024.11.05.622168}, eukaryotes have significantly more disordered regions than bacteria and euryarchaeota: we hence decided to focus on eukaryotes to extract IDRs. We downloaded all 2764 eukaryota proteomes from UniProt and ran Metapredict V3 command metapredict-predict-idrs \cite{emenecker2021metapredict} with default disorder threshold of 0.5 on each one of them. We removed sequences shorter than 30 amino acids and clustered the dataset with mmseqs2 linclust command, with minimum sequence identity threshold of 0.9, 0.8 coverage in coverage mode 1. This pipeline produced an IDP dataset consisting of 30,337,340 sequences.

We fine-tuned ESM-2 models on the IDP-Euka-90 dataset, using a 1\% randomly sampled subset for validation. Fine-tuning was performed on the masked language modeling (MLM) task using four H100 GPUs. We employed a learning rate of 4e-4, consistent with the original ESM pretraining setup. For ESM2-8M, we used a batch size of 700, and for ESM2-150M, a batch size of 96 with 10 gradient accumulation steps. Models were trained for a single epoch to preserve the representations learned during pretraining and avoid overfitting to the downstream dataset.

\section{Geometric feature calculation}\label{app:geo}

We explain here how all geometric features are calculated for a 3D protein structure containing $N$ residues with Cartesian coordinates $\{r_i\}_{i=1}^N$, indexed according to the residue's position in the protein sequence. The features ($R_e$, $R_g$, $\Delta$) are computed separately for each conformation then averaged over the ensemble, whereas the Flory scaling parameters ($\nu$, $A_0$) are fit using the full ensemble (details given below).

As in \cite{dima2004}, we calculate the radius of gyration and asphericity features using the mass-weighted Gyration tensor, $\mathbf{T} \in \mathbb{R}^{3 \times 3}$, computed as
\begin{align}
 T_{\alpha\beta} = \frac{1}{M}\sum_{i=1}^{N}m_{i}\tilde{r}_{i\alpha}\tilde{r}_{i\beta}
\end{align}
where $m_i \in \mathbb{R}$ is the mass of residue $i$, and $\tilde{\mathbf{r}}_i \in \mathbb{R}^3$ are its coordinates after subtracting the center of mass. We denote with $\{\lambda_j\}_{j=1}^3$ the eigenvalues of the gyration tensor $\mathbf{T}$, 

\paragraph{End-to-end distance ($R_e$)} The Euclidean distance between the first and last residue in the sequence:

\begin{align}
    R_e = |r_1 - r_N|
\end{align}

\paragraph{Radius of gyration ($R_g$)}
A geometric property that describes how the protein's mass is distributed about its center of mass - equivalent to the root-mean-square distance of all atoms from the protein's center of mass. It can be calculated using $\mathbf{T}$ as
\begin{align}
    R_g = \sqrt{\text{tr}(\mathbf{T})}
\end{align}

\paragraph{Asphericity ($\Delta$)} Characterises the degree to which a protein's three-dimensional shape deviates from a perfect sphere. Calculated using $\mathbf{T}$ as
\begin{align}
    \Delta = \frac{3}{2} \frac{\sum_{j=1}^{3}(\lambda_j - \bar{\lambda})^2}{(\text{tr}(\mathbf{T}))^2}
\end{align}


\paragraph{Flory scaling exponent and prefactor ($\nu$, $A_0$)} Parametrise the power-scaling-law relationship describing how the Euclidean distance between residues scales as a function of their spacing in sequence. Following the implementation used by \cite{tesei2024idrome}, we fit this relationship to residues spaced at least 5 amino acids apart:

\begin{align}
     |r_i - r_j| = A_0 |i - j|^\nu \hspace{5px} ; \hspace{25px} |i - j| > 5
\end{align}

Unlike the other geometric features which are calculated for each conformation separately and then averaged, the Flory scaling parameters are calculated by first averaging the inter-residue distances observed for each spacing across the whole ensemble, then using the \texttt{optimize.curve\_fit} function provided by SciPy \cite{2020SciPy-NMeth} to fit the ($\nu$, $A_0$) parameters.

\section{Comparisons with existing IDP models}\label{app:IDP}

We evaluate two prominent methods for IDP property prediction: ALBATROSS \cite{lotthammer2024} and STARLING \cite{novak2025}. ALBATROSS is a family of 5 recurrent neural network models, each trained to independently predict one of the ensemble-averaged geometric features $\{R_e, R_g, \Delta, \nu, A_0\}$ directly from sequence. STARLING is a generative diffusion model which generates a conformational ensemble of IDPs by denoising a latent representation of residue-residue distance maps for each conformation. We follow the method used by \cite{novak2025} for property prediction with STARLING: we sample 1000 conformations using 25 DDIM steps, then using the generated ensemble to calculate the ensemble-averaged geometric feature values for each sequence.

We evaluate the publicly released models for both methods on our test set, however we also note that the IDP datasets used
to train ALBATROSS and STARLING notably differ from our training dataset Human--IDRome. In particular, their datasets contain synthetic as well as biological IDP sequences, and the conformational ensembles were generated via coarse-grained MD using an adapted version of the Mpipi force field \cite{joseph2021} rather than CALVADOS-2. We therefore additionally retrained these models from scratch on the Human--IDRome dataset, and report results using both the pretrained and retrained versions of these models in Table~\ref{table:main}.

\subsection{STARLING}
Following the preprocessing in STARLING \cite{novak2025}, we first downsampled the frames for each sequence in our Human--IDRome dataset to reduce the correlation between conformers. We found that keeping every 20th frame was sufficient to stabilise model training, resulting in 50 conformers for each sequence. After downsampling, we used the same hyperparameters and methodology as in \cite{novak2025} to sequentially retrain the STARLING VAE and DDPM models from scratch using our train and validation splits.

\subsection{ALBATROSS}
\subsubsection{Model versions}
In our evaluation of the pretrained ALBATROSS models, we use the default (V2) models available via the SPARROW GitHub repository (\href{https://github.com/idptools/sparrow}{https://github.com/idptools/sparrow}). For predicting $R_g$ and $R_e$ with ALBATROSS, we used the "scaled" versions of these models as recommended.

\begin{table}[t]
\centering
\begin{tabular}{c | c c c c | c} 
 \hline
 
  Predictor &  Number of layers & Hidden size & Learning rate  & Batch size & \# Parameters \\ [0.5ex] 

  \hline
 
 $R_{e}$ & 1 & 55 & 1e-2 & 128 & 34K \\
 
 $R_{g}$ & 1 & 55 & 5e-3 & 128 & 34K \\

 $\Delta$ & 2 & 55 & 1e-2 & 128 & 107K \\

 $\nu$ & 2 & 35 & 5e-3 & 64 & 46K \\
 
 $A_0$ & 1 & 70 & 1e-3 & 128 & 52K \\
  
 \hline
 
 \end{tabular}
\caption{Hyperparameters used for the ALBATROSS (retrained) models.}
\label{table:albatross_hyperparameters}
\end{table}

\subsubsection{Retraining}
We use the same model architecture hyperparameters (number of hidden layers, hidden size) for each feature as used in the published V2 models. We found that we could improve training stability and performance by replacing the loss function used by \cite{lotthammer2024} with the mean of the L1 loss over a batch rather than the sum, and performing a grid search over batch sizes $\{$64, 128, 256$\}$ and learning rates $\{$1e-3, 5e-3, 1e-2$\}$ for each model. We report the best test $R^2$ score achieved over the grid search for each feature in Table~\ref{table:main}, and the hyperparameters used for each model in Table~\ref{table:albatross_hyperparameters}.

\subsubsection{$R^2$ score calculation}
The $R^2$ scores attained in our evaluation of ALBATROSS are notably lower than those reported in the original ALBATROSS work \cite{lotthammer2024}. This discrepancy can be partly explained by a difference in the definition of $R^2$ used between our work and theirs. In \cite{lotthammer2024}, the authors define the $R^2$ score as the square of the Pearson correlation coefficient between the true and predicted values, whereas here we define $R^2$ as the coefficient of determination. 

For targets and predictions $\{(y_i, f_i)\}_{i=1}^{N}$ with target mean $\bar{y} = \frac{1}{N}\sum_{i=1}^{N}y_i$, we calculate the coefficient of determination ($R^2$) as

\begin{align}
    R^2 = 1 - \frac{\sum_{i} (y_i - f_i)^2}{\sum_i (y_i - \bar{y})^2}
\end{align}

which, in general, is lower than the square of the Pearson correlation coefficient - and can even be negative.

\subsubsection{Flory scaling parameters}\label{app:albatross_flory}

We compute the Flory scaling parameters by fitting a power-law relationship between the Euclidean distance of residue pairs and their sequence separation. Following the methodology of the Human-IDRome paper \cite{tesei2024idrome}, we exclude residue pairs with a sequence separation of less than five, as these short-range interactions are governed by local chain stiffness and deviate from the global scaling law, which may differ from that used by ALBATROSS \cite{lotthammer2024}.

\end{document}